\documentclass[runningheads]{llncs}
\usepackage{graphicx}
\usepackage[graphicx]{realboxes}
\usepackage{rotating}
\usepackage{capt-of}

\begin{document}
%
\title{A Semi-automatic Data Extraction System for Heterogeneous Data Sources: A Case Study from Cotton Industry}
\titlerunning{A Semi-automatic Data Extraction System for Heterogeneous Data Sources}
%
%
\author
{
Richi Nayak\inst{1}\and
Thirunavukarasu Balasubramaniam\inst{1}\and
Sangeetha Kutty\inst{1}\and
Sachindra Banduthilaka \inst{2}\and
Erin Peterson \inst{3}
}
\authorrunning{Richi Nayak et al.}

\institute{School of Computer Science and Centre for Data Science,
Queensland University of Technology,
Brisbane, Australia. \email{\{r.nayak,thirunavukarasu.balas,s.kutty\}@qut.edu.au}  \and
Redeye Apps Pvt Ltd, Brisbane, Australia. \\\email{sachi.banduthilaka@redeye.co} \and
Erin Peterson Consulting, Brisbane, Australia.\\ \email{erin@peterson-consulting.com}
\\
}
\maketitle              
\begin{abstract}
With the recent developments in digitisation, there are increasing number of documents available online. There are several information extraction tools that are available to extract information from digitised documents. However, identifying precise answers to a given query is often a challenging task especially if the data source where the relevant information resides is unknown. This situation becomes more complex when the data source is available in multiple formats such as PDF, table and html. In this paper, we propose a novel data extraction system to discover relevant and focused information from diverse unstructured data sources based on text mining approaches. We perform a qualitative analysis to evaluate the proposed system and its suitability and adaptability using cotton industry.

\end{abstract}
\section{Introduction}

In the recent years, agricultural data is available in a wide variety of sources. Information overload digitising the documents has resulted in considerable benefits such as accessibility, cost savings and disaster recovery. Due to these advantages many enterprises digitise their documents online resulting in abundant data on the web and with the enterprises~\cite{krallinger2017information}. The digitization of newspaper articles, government policy documents, research publications and statistical reports has contributed to the creation of big data~\cite{banawan2018capacity}. One of the challenges of the big data is the information overload due to which it is difficult for humans to sift through the information on a subject topic and identify the relevant value for the information of interest. For example, it is a challenge for a farmer or a policy officer in an agriculture firm to obtain an exact value or a trend on a specify query "Carbon footprint per bale" from the vast amounts of related reports available online. Often the documents containing the term will be returned and then a manual search on the returned documents is required to identify the relevant
value for the query of interest.

Information retrieval techniques have been applied in various scenarios to retrieve relevant information for users from a data collection based on user queries~\cite{croft2010search,mayer2013big}. A user query is typically a short text with keywords representing the user requirement. Based on the given query, most relevant information is retrieved from the collection of data~\cite{wu2014data}. Majority of search engines including Google can only retrieve a list of documents that match the query, but they cannot return a value for the given query.

Corporations are under increasing pressure to produce annual reports describing their environmental, social, and economic sustainability~\cite{porter2006link}. These corporations set sustainability targets (e.g. reduce energy usage by 10\%) and use data-derived indicators to measure progress towards those goals (e.g. total energy used in production). However, the data needed to generate these sustainability indicators are often stored in a number of disparate data sources~\cite{peterson2017assessment}.To obtain domain-specific indicators, a manual search of information (i.e. values for indicators) from the data collection is the obvious choice. It is a cumbersome process to manually search for these indicator values using a search engine and then read through all relevant documents to identify the exact answer. An automated method of obtaining these domain-specific indicator values will be desirable.

With the recent advancements in automation, it is possible to semi automatically generate annual reports for companies that will contain information about data-derived indicators and targets set by the organization. This report will propose a system that could collect information on the indicators from various sources and, then synthesize the collected information to generate a report semi-automatically.

In this paper, we present a novel system that has been used in the Australian Cotton industry to help the users to sift through the Big data collected from various sources such as websites, published articles, case studies, reports and identify the value of the domain-specific indicators automatically. The methodology used in this system based on text mining approaches is used to scan and filter the relevant information from different unstructured data sources~\cite{banawan2018capacity}. This information will then be collected to create a repository for inquiry and automatically generating a report with relevant values for the query of interest.

\section{Background and Related Work}
\textbf{Content extraction from web pages:} The data in web pages are commonly represented as HTML and PDF documents where extraction of relevant content from these documents is essential to perform a text mining task~\cite{gupta2003dom}. Extraction of relevant content from HTML page is straightforward as the HTML documents are annotated with structured tags~\cite{lin2002discovering}. Mostly $<$p$>$, $<$a$>$ and $<$table$>$ tags cover the important information. While, it is easy to extract content in $<$p$>$ and $<$a$>$, extracting meaningful information from tables is challenging. A heuristic approach can be used to extract table information in a cell format that makes the information retrieval easy~\cite{wei2006table}. Extracting content from PDF documents is challenging. The Optical Character Recognition (OCR) approach is widely used to read the contents of a PDF file in text format. 

\textbf{Chunking (Shallow parsing) and Named Entity Recognition (NER):} Text mining techniques such as chunking and NER help to retrieve any phrase from the document that follows a regular expression and to identify the named entities in the document respectively. In chunking, a sentence or text is analysed by dividing it into syntactically related non-overlapping groups of words such as phrases~\cite{sang2003introduction}. NER is the task of identifying named entities (phrases that contains the names of persons, organizations or locations) in a text/sentence. Traditional approaches use sequential models like HMM (Hidden Markov Model) and CRF (Conditional Random fields) reply on handcrafted features~\cite{finkel2005incorporating}. Whereas the emerging approaches use neural network models such as RNN/LSTM with word embedding~\cite{zhai2017neural,habibi2017deep}. For a sentence that doesn’t follow the given regular expression, Parts of speech (POS) can be used~\cite{grishman1997information}. POS helps to focus only on the set of meaningful phrases in the documents and querying this set will return more relevant results. NER is necessary to identify the named entities in the document that can reveal the spatio-temporal information which is vital in many cases~\cite{strotgen2010extraction}. A simple spatio-temporal query needs spatio-temporal information to be known during the query and it is not possible to automatically reveal the spatio-temporal nature of the documents and hence a proper integration of Chunking and NER is essential to retrieve any results with underlying spatio-temporal information captured during the query process.

\textbf{Indexing the dataset:} Indexing has been widely utilized to facilitate query searches~\cite{kononenko2014mining}. Elasticsearch~\cite{akdogan2015elasticsearch} is a popular document-oriented search engine that stores the data as an information repository. It provides the environment to deal with Big Data processing. Unlike traditional database systems, data is partitioned and stored as a collection of documents in Elasticsearch. It also maps the data using a dynamic mapping that makes it searchable using TF-IDF (Term-Frequency x Inverse Document Frequency) and BM25 methods~\cite{ramos2003using,perez2009integrating}.

Though each of these concepts has been used distinctively in various problems, the capability of these methods together is not well studied. In this paper, we integrate these concepts together to facilitate the framework for automated discovery of relevant information from the unstructured web data sources. We first extract contents from web pages and process various types of data. We perform chunking and NER on the pre-processed data that is subsequently stored and indexed in the Elasticsearch search engine. Finally, we perform query redefinition to fine tune to results.

\section{Automated Discovery of Relevant Information}
This section discusses about the five phases involved in the automated discovery of relevant information.
\begin{enumerate}
    \item Data acquisition
    \item Semantic enrichment
    \item Data deposition
    \item Query information
    \item Visualization
\end{enumerate}

\begin{figure}
    \centering
    \includegraphics[width = 12cm, height = 6cm]{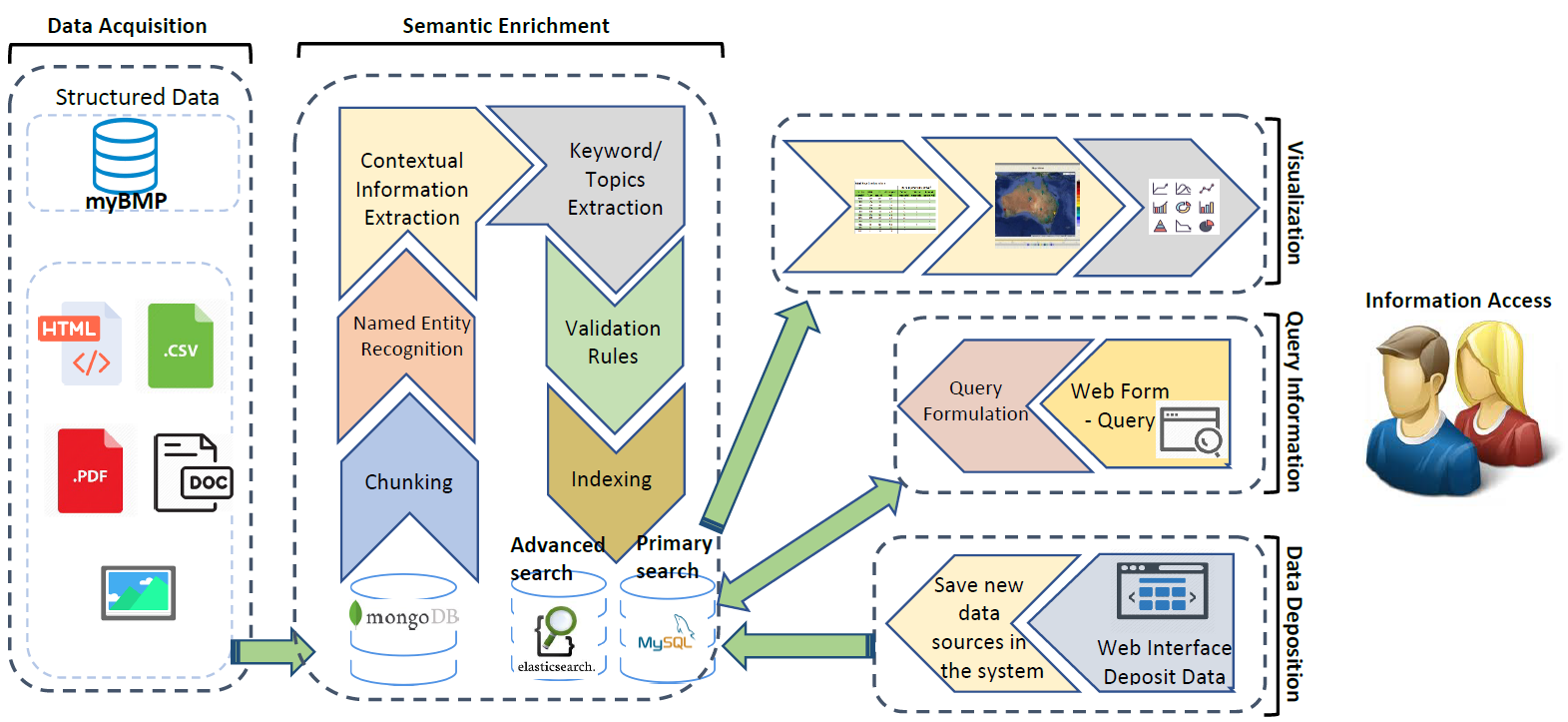}
    \caption{The proposed framework}
    \label{fig:arch}
\end{figure}

We integrate text mining concepts to propose a novel methodology to automatically discover relevant focused information from the unstructured web data sources. As shown in Figure~\ref{fig:arch}, the proposed framework includes three main phases namely: data acquisition, semantic enrichment, and query formulation.

\subsection{Data Acquisition}
In this project, we use several types of data sources such as grey literature, web pages, and databases. Grey literature is often produced for internal communication or for public distribution by government agencies, professional organizations, universities, corporations, research centers, associations, and societies. This type of literature is not available through the usual bibliographic sources such as databases or indexes. Grey literature can be both in print and, increasingly due to digitization, it is in electronic formats as PDF, text and html pages. We envision using technical reports, census reports, survey reports, annual reports, journal articles and conference articles according to their availability and applicability.

In spite of the availability of several sources of data, regrettably, an increased quantity of data does not produce more value; only more of the relevant data will enable us to identify the indicator values correctly and autonomously. The next step is to clean the data for quality outcomes.

\subsubsection{Data deposition}
Although this is one of the major steps, this is an often-neglected step. This step helps to improve the quality of the data that will eventually be used for knowledge discovery. Data gathering methods are often loosely controlled hence resulting in incorrect values and nearly impossible values too (e.g., age: 150 or male: yes and pregnant: true). Also, it is essential not to blindly remove noise as it can result in valuable information being unintentionally eliminated. It requires domain knowledge to remove the incorrect values with a correct definition of “noise”. Otherwise, it can become cumbersome because pre-processing is reliant on the task at hand. 

Text mining models require inputs to be presented in vector format. The transformation from the raw text data to a vector format requires several pre-processing steps such as parsing to remove unwanted tags and texts, stop words removal, stemming, term weighting generation and dimensionality reduction. We explain these steps below.

\paragraph{Data Parsing for different formats:}
Each of the data sources differs significantly in their structure and type of data that it holds.  Hence, it is essential to deal with each of them using different techniques and identify relevant information that will be beneficial to users. The common data sources exist in the formats of HTML, Tables and PDFs. The following steps are required to convert these formats into text format so the querying process can begin to obtain the desired indicator value.

\texttt{HTML Page:} Web pages usually encode the data in a structured template that is suitable for human reading. It has to be converted in a format that is suitable for machines to process. Web scraping involves extracting text from structured documents like HTML, which uses tags that indicate the nature of data. For example, URLs are enclosed within $<$a$>$ tag and paragraphs are enclosed within $<$p$>$. These tag structures are used to identify various elements within HTML documents. We identify all the elements containing text content and scrape them as a set of sentences in a text file.

\begin{figure}
    \centering
    \fbox{\includegraphics[width = 9cm]{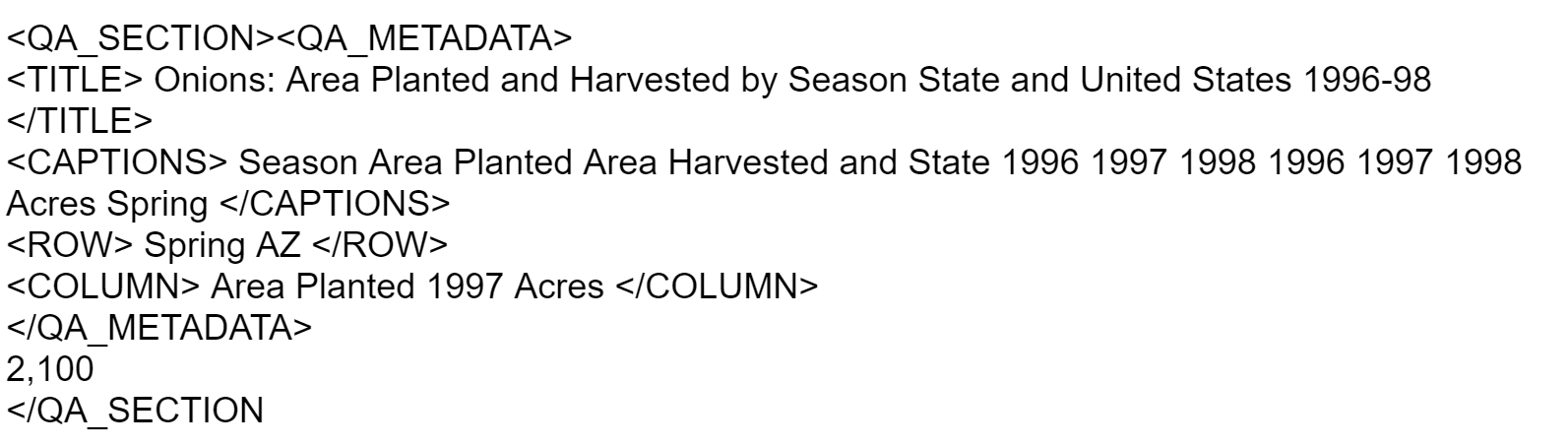}}
    \caption{Original Table}
    \label{fig2}
\end{figure}

\begin{figure}
    \centering
    \fbox{\includegraphics[width = 9cm]{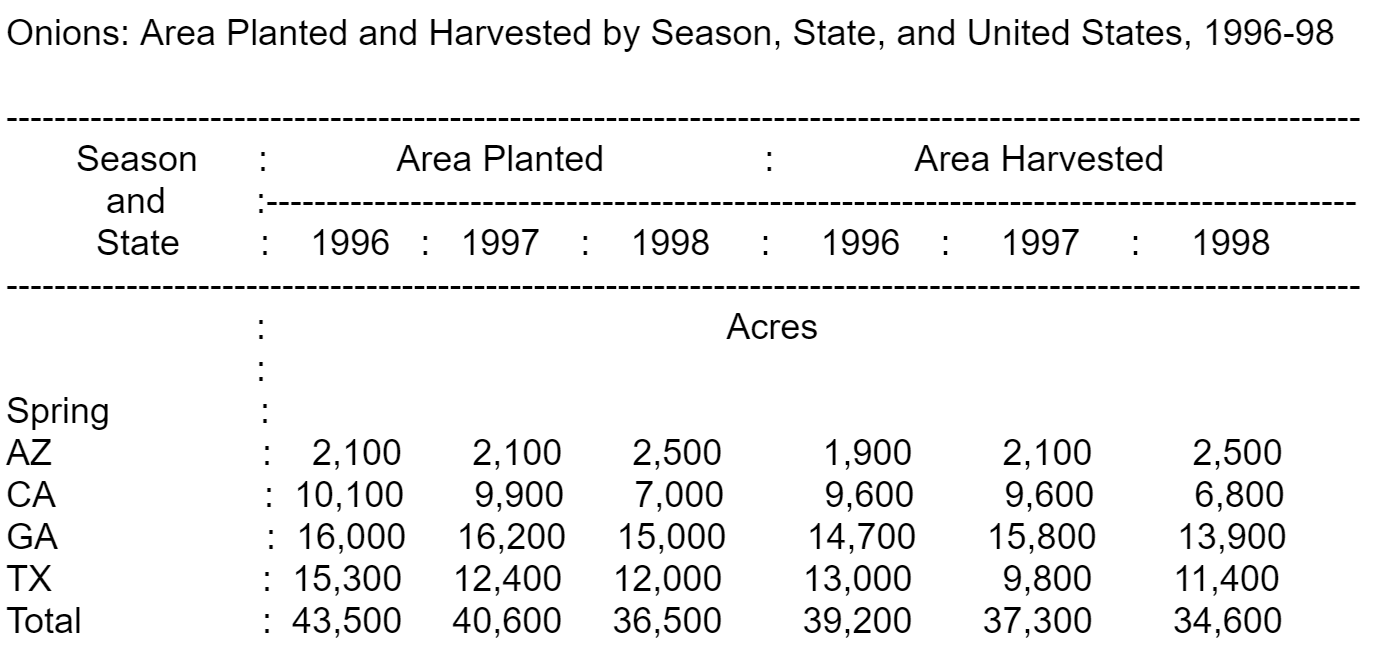}}
    \caption{Extracted Table Headers}
    \label{fig3}
\end{figure}

\texttt{Tables:} The simple scraping of text from table elements (tags) is not suitable for processing as tables usually contain rich information and complex patterns with varying row and column representations. Tables often link their header rows and columns to cells that contain values representing header links as shown in Fig.~\ref{fig2}. To extract meaningful information from a table, a machine learning algorithm called as Conditional Random Fields (CRF) has been applied. First it learns the column and row headers and then labels each line of the table to a header label. Using this approach, we achieve a more realistic data format of the table that is suitable for machines to process as shown in Fig.~\ref{fig3}.

\texttt{PDFs:} Most of the data sources are reports and surveys, which are usually available as PDF files. Unlike HTML pages, PDF files do not have any structural information available as the metadata. This makes the analysis harder. We apply a PDF parsing technique that extracts only the text without any structural information.  PDF parser extracts all the text represented as ASCII code. As PDFs are in human readable form, hence it does not have any structural information for machine to perform tag based extraction. Hence text locations, font sizes and writing directions are used for focused parsing. 

\subsection{Semantic Enrichment using Chunking and NER}
The content, extracted from various sources like HTML pages, PDFs, and tables, is stored in a text file. We assume that the indicator values and associated information (i.e., respective value with units) can be identified correctly if chunking is done. Chunking is a process of identifying a group of chunks/information from a sentence based on Parts of Speech (POS). For each sentence, the words are tagged with their POS and Regular Expressions are used to identify the needed chunks. We assume that any indicators mentioned in a source have a high probability of being followed by its associated value and unit. The regular expressions are defined such that they can extract INDICATOR, VALUE and UNIT. For example, given the extracted text, “The average hectares planted per participant increased slightly 1,518 hectares.”, the following information is identified:
\begin{enumerate}
    \item Possible indicator: planted per participant
    \item Possible value: 1518
    \item Possible unit: hectares
\end{enumerate}

\begin{figure}
    \centering
    \includegraphics[width = 12cm]{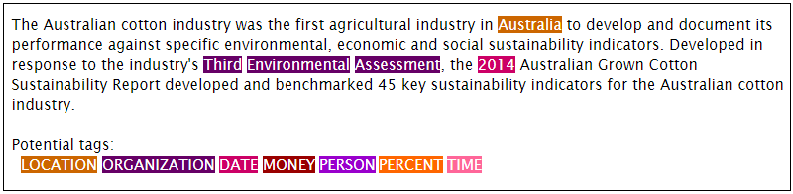}
    \caption{NER Example}
    \label{fig4}
\end{figure}

The data sources contain various levels of spatial and temporal information. It is difficult for a simple querying process to identify them. We highlight the spatial and temporal information from a source using NER. It generates annotated tokens for entities in the text and highlights the names for each entity. NER has near-human performance in identifying the following entities in the English language, Location (Spatial), Organization, Date (Temporal), Money, Person, Percent and Time (Temporal). An example is shown in Fig.~\ref{fig4} where NER identifies two entities with spatial and temporal information. With the annotated tokens for each sentence in a source, we can focus on obtaining only the entities that we are interested.

\paragraph{Indexing:}
\begin{figure}
    \centering
    \includegraphics[width = 7cm]{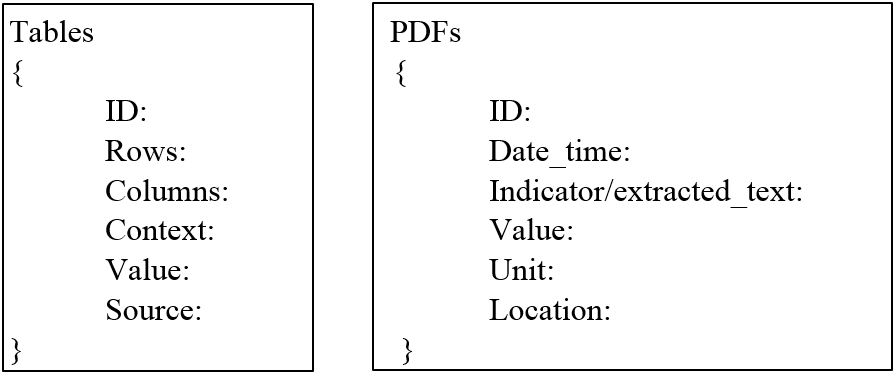}
    \caption{The Mapping Process after parsing the information source}
    \label{fig5}
\end{figure}
Once all the data sources are collected, pre-processed and parsed, we need to store the data and search the relevant information within the repository. Indexing facilitates storing and allows a search engine to retrieve the data quickly. The indexing will optimize and increase the runtime performance; thus finding relevant information from a large data collection can be achieved within a few milliseconds. As we have a distinct parsing technique for a Table source or a PDF source, the data sources of Table and PDF formats are indexed with the mapping process after parsing as shown in Fig.~\ref{fig5}.

\subsection{Query Formulation}
With the information indexed, a query can be posed to the search engine based on the mapping structure to retrieve the information about indicators. We start with simple queries which directly searches for indicators without any modifications. For example, if the value of the indicator “area of cotton planted” is sought, then these four terms are formed as the query. Based on preliminary analyses, it was determined that the queries had to be modified or redefined in some cases to achieve satisfactory outcomes. We modified the query by adding more information. The additional information that found useful in extracting relevant results included
\begin{enumerate}
    \item Simple Query
    \item Simple Query + Keywords
    \item Simple Query + Keywords  + Source
\end{enumerate}

\textbf{Keywords.} Keywords are usually terms that help to easily identify the indicators. We found that indicator unit made it easy to identify the indicator with high precision. Several keywords can be added to a simple query based on the domain knowledge. 

\textbf{Sources.} Sometimes, even with the keywords provided, the query does not provide relevant results. If we know the data source from which the particular indicator can be identified, we can search the query for that particular source alone. By doing this one can avoid searching a big data source and thus any irrelevant information from unwanted sources can be ignored.

\section{Empirical Analysis}
\subsection{Data sources}
The Cotton Research and Development Corporation (CRDC) is required to report multiple aspects of sustainability. However, the data sources where this information appears are de-centralised and are difficult to identify, and those that were identifiable were often appear in PDF format. This poses a significant challenge to CRDC to obtain relevant indicators values on repetitive basis. As a result, we recommend to use the proposed methodology to retrieve cotton-related data from a wide various sources in autonomous fashion.

There were a total of 102 sustainability indicators for which the information was required. Some example indicators inlcude: ``Area of cotton planted"; ``Irrigated planted area"; ``Dryland planted area"; and ``Gross value cotton lint". The possible data sources where this information can appear were obtained after consultation with domain experts. Some of the sources where the data was obtained are: Cotton Australia Website; Cotton Grower Yearbook 2016; Cotton Grower Surveys 2011, 2013 and 2018; Annual agricultural-commodities reports; Australia cotton shippers; and the Australian Bureau of Statistics (ABS) Website. There were several static data sources such as journal articles and technical reports. 

\subsection{Implementation}
Open-source tools such as Python, MongoDB and Elasticsearch have been used to implement the proposed methodology. The unstructured data from various sources are parsed using Python and stored in a MongoDB database. The pre-processing, chunking and NER, as implemented in Python moduels, will decompose the extracted text from MongoDB into classes as shown in Fig.~\ref{fig5}, which is then indexed in a local Elasticsearch server. The queries are fed into an Elasticsearch search engine to retrieve relevant indicator values. A simple user interface is built for user interaction.

\subsection{Results: Information Retrieval Performance}
Once the data sources were identified, we applied the proposed methodology to each indicator to obtain the results. We also assess how suitable and adaptable is a data source and the indicator for autonomous sustainability reporting purpose. Based on the categorization of each indicator, we identify if the particular indicator is adaptable.

Table~1 summarizes the findings for a few sample indicators and Table~\ref{table_or} details the overall results. A total of 102 indicators deemed to retrieve high performance (i.e., high relevance score). A total of 15 indicators had moderate values; however, we were unable to locate data for 23 indicators. The reason may be that data may not exist for these indicators and need specific help to collect the possible data sources where the values may reside. We plan to implement some sophisticated machine learning methods to identify values of these indicators if they exist in collected sources. We also plan to have workshops with CRDC staff to get more information on data sources where this information may appear.
\begin{table}
\label{resultsum}
\caption{Result summary}
\begin{sideways}

\begin{tabular}{|c|p{3cm}|p{3.5cm}|p{1cm}|p{1.5cm}|p{1.5cm}|c|c|p{1.5cm}|p{1.5cm}|}
\hline
S.No & Indicator & Query & Data source & Source Type & Added Keywords & Suitability & Adaptability & Relevance score & Result achieved \\
\hline
1 & Area of cotton planted & Cotton area planted 2016 & D1 & Table & ha & H & H & 0.59 & Y \\
2 & Irrigated planted area & Irrigated planted area & D2 & PDF & ha & M & L & 0.30 & Y \\
3 & Cotton exports & Export million tonnes & D3 & PDF & Million tonnes & M & M & 0.40 & Y \\
4 & Total amount of cotton produced & Total amount of cotton produced (metric tonnes OR million bales) & D1 & HTML & Metric tonnes OR million bales & H & H & 0.73 & Y \\
5 & Percentage change: Crop rotation & Crop rotation & D4 & PDF & \% & L & M & 0.63 & Y \\
6 & Native vegetation area per farm & Native vegetation area per farm & D4 & HTML & Percentage farm land & M & H & 0.48 & Y \\
7 & Regional gross production value & Regional gross production value & D3 & PDF & \% & M & M & 0.28 & Relevant results \\
8 & Average cotton area per farm & Average cotton area per farm & D1 & Table & ha/farm & M & H & 0.22 & Y \\
9 & salinity & salinity & D4 & PDF & mm/year & L & L & 0.00 & N \\
10 & Percentage of growers using integrated pest management & Percentage of growers using integrated pest management & D5 & PDF & \% & H & H & 0.63 & Y \\
11 & Irrigation water use index & Water use on Australian farms cotton & D6 & HTML & \% & M & H & 0.24 & Y \\
12 & Groundwater levels & Groundwater levels & D7 & PDF & million tonnes & H & H & 0.69 & Y \\
13 & Workers receiving regular health and safety training & Workers receiving regular health and safety training & D4 & PDF & \% respondents or \% & M & M & 0.35 & Y \\
\hline
\end{tabular}

\end{sideways}
\end{table}

\begin{table}[tb!]
\caption{Overall results}
\label{table_or}
\centering
\begin{tabular}{|c|c|c|c|c|}
\hline
Data Type & Total Queries & Results achieved & Relevant results & Results not achieved\\
\hline
HTML & 13 & 10 & 3 & 0 \\
PDF & 76 & 51 & 12 & 13 \\
Table & 3 & 3 & 0 & 0\\
Unknown & 10 & 0 & 0 &  10 \\
\hline
Total & 102 & 64 & 15 &  23 \\
\hline
\end{tabular}
\end{table}
For every indicator, we analyze the query results in depth to identify Suitability and Adaptability of the current sources in the repository. 

\textbf{Suitability:} We measure the suitability to identify if the existing data is suitable for the development of the Sustainability report. Based on the query analysis, we categorize suitability as High, Medium and Low.

Each query returns a list of matching results. Each returned result is ranked to identify the best match. We use the normalized relevance score to decide the suitability category of the indicators. The normalized relevance score is to measure how relevant is the results retrieved for the given query. The relevance score is calculated using term frequency $\times$ inverse document frequency, and field length norm. These measures are combined and normalized for each source type to rank the indicators. Higher the value, the higher is the suitability.

\begin{table}[tb!]
\caption{Suitability score}
\label{table_suitability}
\centering
\begin{tabular}{|c|c|}
\hline
Rank level & Score \\
\hline
High & 0.7 to 1  \\
Medium & 0.4 to 0.6 \\
Low & 0 to 0.3 \\
\hline
\end{tabular}
\end{table}

\textbf{Adaptability:} We use adaptability to identify the continuous (ongoing) extraction capacity of the indicators and the sources. We categorize adaptability using two factors as shown in Table~\ref{table_adaptability}. The adaptability score is an integrated score derived from two scores related to query and data dependencies as shown in Table~\ref{table_adaptabilitys}.
\begin{table}[tb!]
\caption{Adaptability Categories}
\label{table_adaptability}
\centering
\begin{tabular}{|c|c|p{5cm}|}
\hline
Rank level & Query-dependent & Data-dependent \\
\hline
High & 2 or more query re-definitions & Source subscription and source-specific search  \\
Medium & 1 query re-definition & Source specific search \\
Low & No query redefinition & Open source\\
\hline
\end{tabular}
\end{table}

\begin{table}[tb!]
\caption{Adaptability score}
\label{table_adaptabilitys}
\centering
\begin{tabular}{|c|c|c|}
\hline
Query-dependent & Data-dependent & Adaptability score \\
\hline
L/M & L & H  \\
L/M & M & M \\
L/M & H & L \\
H/M & L & M \\
H/M & M/H & L \\
\hline
\end{tabular}
\end{table}

\textbf{Query-dependent:} Any indicator that is query dependent is considered as medium/high category. Query-dependent is a condition when a simple query with keywords does not result in a good retrieval. Sometimes, the indicators are represented in different terms in the sustainability report and the data source. The situation is similar when the source uses different units to measure the same indicator. By carefully redefining the query with the possible alternative terms and units, this can be avoided to get better query results. Out of 102 indicators, 66 indicators are retrieved with high performance using units as keyword whereas only 48 indicators can be retrieved without keywords. We categorize the query-dependent as high, medium and low based on the level of query redefinition required (Table~\ref{table_adaptabilitys}).

\textbf{Data-dependent:} Any indicator that is data dependent is considered as a medium/high category. Data-dependence is a condition when you need a subscription to access the source data or when you do a source-specific search. Even though source specific searches will not affect in retrieving relevant results, in future if that specific source is not available, the indicator may be difficult to query. Table~\ref{table_queryredef} shows an example of 3 indicators where the importance of Keywords and Data source in retrieving relevant results are analyzed. While "Irrigated planted area" can return the required result without any query redefinition, "Cotton exports" needs "Million tonnes" as a keyword to achieve the result. On the other hand, indicator "Cotton stubble" is data dependent and results may not be achieved if that specific data source (D4) is not available in the repository. 

\begin{table}[tb!]
\caption{Query Redefinition}
\label{table_queryredef}
\centering
\begin{tabular}{|p{2.2cm}|p{2.2cm}|p{2.4cm}|p{1.9cm}|p{1.2cm}|p{1.4cm}|}
\hline
Indicator & Simple query & Keyword (Query dependent) & Source (Data dependent) & Result achieved & Relevance score\\
\hline
Irrigated planted area & Irrigated planted area &  &  & Y & 13.35 \\
\hline
Cotton exports & Cotton exports &  &  & N & 9.67 \\
Cotton exports & Cotton exports & Million tonnes &  & Y & \textbf{17.57} \\
\hline
Cotton stubble & Cotton stubble &  &  & N & 12.34 \\
Cotton stubble & Cotton stubble & \% &  & N & 12.34 \\
Cotton stubble & Cotton stubble & \% & D4 & Y & \textbf{17.03} \\
\hline
\end{tabular}
\end{table}



\section{Conclusion}
With the growing advances in the web and the collection of unstructured data, it is crucial to process and analyze them for various purpose. Especially when the data sources have textual information in the different forms, the retrieval of relevant information becomes challenging. In this paper, we propose a methodology to automatically discover the relevant information from the collection of unstructured web data sources. We combine several text mining techniques to extract, store and perform a query search.

We implemented this methodology for a corporation for their specific need of obtaining information on sustainability indicators autonomously. We presented an overall results as well as we provided a qualitative analysis detailing the suit- ability and adaptability of the query and data sources respectively. The proposed method enables the corporation to utilise the Information Repository built from a wide variety of sources for interrogating the relevant information for sustainability reporting.

The proposed method can be applied to similar problems. In future, we will adopt a learning-to-rank concept in query redefinition that eases the human input. 
\bibliographystyle{splncs03_unsrt}
\bibliography{air}
\end{document}